\documentclass[aps,pra,superscriptaddress,amsmath,amssymb,preprintnumbers,floatfix,showpacs,11pt]{revtex4}
\usepackage{amssymb}
\usepackage{epsfig}
\usepackage{subfigure}
\usepackage{graphicx}
\begin{document}
\title{Comment on "Deterministic six states protocol for quantum communication" [Phys. Lett. A 358
(2006) 85]
 }
\author{ Faisal A. A. El-Orany }
\email{el_orany@hotmail.com, Tel:006-166206010, Fax:
0060386579404}
 \affiliation{ Cyberspace Security
Laboratory, MIMOS Berhad, Technology Park Malaysia, 57000 Kuala
Lumpur, Malaysia}

\date{\today}

\begin{abstract}
In [J.S. Shaari, M. Lucamarini, M.R.B. Wahiddin, Phys. Lett. A 358
(2006) 85-90] the deterministic six states protocol (6DP) for
quantum communication
 has been  developed. This protocol is based on  three mutually unbiased bases and
four encoding operators. Information is transmitted between the
users via two qubits from different bases. Three attacks have been
studied; namely intercept-resend attack (IRA), double-CNOT attack
(2CNOTA) and quantum man-in-the-middle attack. In this Letter, we
show that the IRA and 2CNOTA are not properly addressed. For
instance, we show that the probability of detecting Eve in the
control mode of
 the IRA  is $70 \%$ instead of $50\%$ in the previous study.
Moreover, in the 2CNOTA, Eve can only obtain  $50\%$ of the data
not all of it as argued earlier.

\end{abstract}

 \pacs{03.67.Dd, 03.67.Hk}
   \maketitle

   {\bf Key words:} quantum protocol, deterministic protocol,
   six states protocol, eavesdropper, intercept-resend attack, double-CNOT attack

\vspace{1cm}
 In \cite{shaari}, the authors  have developed the
deterministic six states protocol (6DP) for quantum communication.
This protocol can be briefly explained as follows: We have three
bases, namely, $x,y,z$, which are mutually unbiased and have the
forms:
\begin{eqnarray}
\begin{array}{lr}
  |z_+\rangle=|0\rangle, \quad |z_-\rangle=|1\rangle,\\
  |x_\pm\rangle=(|z_+\rangle\pm |z_+\rangle)/\sqrt{2}, \quad
   |y_\pm\rangle=(|z_+\rangle\pm i|z_+\rangle)/\sqrt{2}.
   \label{1a}
\end{array}
\end{eqnarray}
Bob chooses  two qubits  from (\ref{1a}) {\it on condition that
they are from different bases} and send them to Alice.
 Alice upon receiving the qubits  operates
on both of them with one of the four operators $\hat{I} , \hat{X},
i\hat{Y}$ or $\hat{Z}$ according to the message she wants to
address to Bob. Then she sends them back  to Bob. He then measures
the qubits in the   bases,  which they have been prepared earlier,
to decode the key. The key extraction  depends on an advance
agreement between the users related to the encoding processes,
i.e. $\hat{I}\longrightarrow 00, \hat{X}\longrightarrow 10,
i\hat{Y}\longrightarrow 01,\hat{Z}\longrightarrow 11$. This
scenario   makes  the 6DP  more secure than the standard BB84
protocol \cite{Bennett}. As we can see, no public discussion is
necessary for the completion of the encoding-decoding procedure,
as it happens in BB84.
 Moreover, the BB84 protocol has a
probability of $1/2$ that a transmitted bit has to be discarded
due to the wrong choice of basis on both sides. This is not the
case in 6DP, where one of the users knows the measurement bases.
We proceed, three types of attacks, namely, intercept-resend
attack (IRA), two control-NOT gates \cite{Nielsen}
 attack (2CNOTA)   and quantum
man-in-the-middle attack, have been discussed for 6DP in
\cite{shaari}. It has been shown that  Eve can obtain $50\%$ of
the message by using IRA. Also, she can obtain   full information
from the protocol  via 2CNOTA. By means of the quantum
man-in-the-middle attack Eve either induces two errors in the
communication, one for each path, or induces no errors at all.
This is something that a "natural" noise cannot statistically
give. This represents a sure signature of Eve's presence [1].

In \cite{shaari}, the treatments of the   IRA and the 2CNOTA have
been performed when the users use two  qubits from the same bases
(, precisely, identical qubits). This contradicts with the notion
of the protocol, which is based on two qubits from different
bases. Actually, this property is necessary to  establish  the key
in the protocol.
 For instance, assume that Bob sends one particle in the state, say, $|x_+\rangle$ and
after a complete round he receives the particle in the same state.
For him, this means that Alice used $\hat{I}$ or $ \hat{X}$ as an
encoder and hence the key cannot be extracted.
 As a result of this,  we deal again with these two attacks
  considering two qubits from different bases, as it should
be. In treating IRA, we consider the permutation of different
bases in the protocol. We  obtain results  completely different
from the previous ones [1].   For instance, we show that the
probability of detecting Eve in the control mode of
 the IRA  is $70 \%$ instead of $50\%$ in [1]. Additionally, in contrast with the
analysis given in [1],
  we show that 2CNOTA is not relevant to be used
against this protocol.  Throughout the investigation we use the
same notations and some of the equations given in \cite{shaari}.

We start the analysis with the IRA for control mode. This  can be
explained as follows:
 Bob
has to prepare two states, say, $|B_1,B_2\rangle$
  from different bases and sends them
to Alice, who returns them back to Bob without doing any action on
them. Bob checks if Eve is on  line or not by measuring these
qubits using the same  bases in which they have been initially
prepared. In IRA, Eve measures both of the travelling qubits in
the forward and in the backward paths in the same bases. Suppose
that Eve performs projective measurements along the orthogonal
bases \cite{shaari}:
\begin{equation}\label{1b}
    |\chi_j\rangle=\cos(\frac{\theta_j}{2})|0\rangle+\exp(i\varphi_j)
    \sin(\frac{\theta_j}{2})|1\rangle,\quad
      |\chi_j^\bot\rangle=\exp(-i\varphi_j)\sin(\frac{\theta_j}{2})
      |0\rangle-\cos(\frac{\theta_j}{2})|1\rangle,
      \quad j=1,2,
\end{equation}
where $0\leq\theta_j\leq\pi, 0\leq \varphi_j\leq 2\pi$. The index
$j$ indicates that Eve can use  one or more different bases for
the measurement.  In IRA, the probability that Alice (Bob) does
not detect Eve on the forward (backward) path equals the
probability to obtain the states $|B_1,B_2\rangle$ prepared
initially by Bob. These two partial probabilities are equal and
each of which can be expressed as:
\begin{equation}\label{1cref}
P^{|B_1,B_2\rangle}_{noEve_A}=P^{|B_1,B_2\rangle}_{noEve_B}=P_{B_1|\chi_1}P_{B_2|\chi_2}
+P_{B_1|\chi_1^\bot}P_{B_2|\chi_2^\bot},
\end{equation}
where $P_{B_j|\chi_{j'}}=|\langle B_j|\chi_{j'}\rangle|^2,
j,j'=1,2$ and the subscripts $noEve_A$ and $noEve_B$ mean that the
probabilities are related to Alice (forward path) and Bob
(backward path), respectively.
\begin{figure}
\includegraphics[width=.60\linewidth]{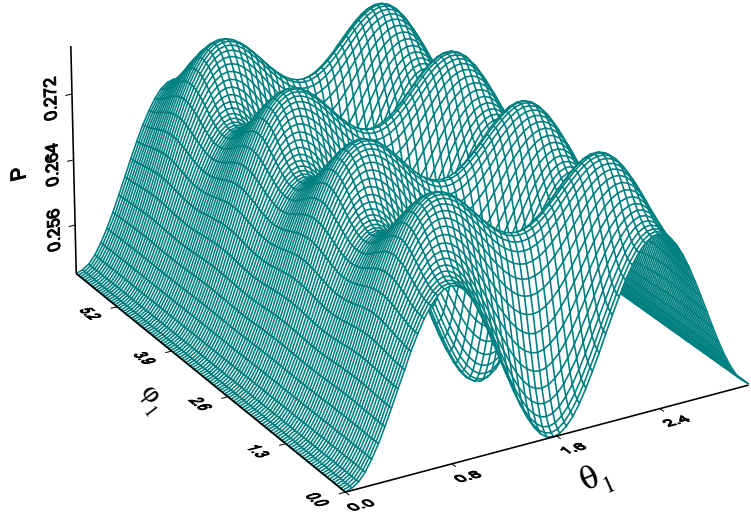}
   \caption{ The average probability of not detecting  Eve  during
the control mode of the  IRA. The angles $\theta_1$ and
$\varphi_1$ define Eve's projective measurement direction. The
angles are measured in radian. }
\end{figure}
  The probability $P^{|B_1,B_2\rangle}_{noEve}$
 that Eve is not detected after a
whole control run is  the product of the two partial
probabilities, i.e
$P^{|B_1,B_2\rangle}_{noEve}=P^{|B_1,B_2\rangle}_{noEve_B}P^{|B_1,B_2\rangle}_{noEve_A}$.
This quantity  is quite different from Eq. (5) in [1].
 Now we  study the behavior of $P^{|B_1,B_2\rangle}_{noEve}$ for two cases  depending on whether Eve measures
the travelling qubits (, i.e., qubits in the transit) using the
projective qubits from the same or different bases. The first case
(, i.e. $j=1$ in (\ref{1b})) is given for the sake of comparison
with the technique used in [1]. Of course this scenario  gives Eve
complete information on one of the qubits and probabilistic
information on the other. In spite of this, the results will be
sufficient and better than the earlier ones [1]. Starting with the
first case, from (\ref{1a}) and (\ref{1b}) the quantities
$P_{B_j|\chi_1}$ (where $\chi_1=\chi_2$) can be easily calculated
and the average probability of not detecting Eve $P$ in the IRA
is:
\begin{eqnarray}
\begin{array}{lr}
P=\frac{1}{12}\left(P^{|z_+,x_+\rangle}_{noEve}+P^{|z_+,x_-\rangle}_{noEve}+
P^{|z_-,x_+\rangle}_{noEve}+P^{|z_-,x_-\rangle}_{noEve}+
P^{|z_+,y_+\rangle}_{noEve}+P^{|z_+,y_-\rangle}_{noEve}\right.\\
\\
\left. + P^{|z_-,y_+\rangle}_{noEve}+P^{|z_-,y_-\rangle}_{noEve}
+P^{|y_+,x_+\rangle}_{noEve}+P^{|y_+,x_-\rangle}_{noEve}+
P^{|y_-,x_+\rangle}_{noEve}+P^{|y_-,x_-\rangle}_{noEve}\right).
\label{1d}
\end{array}
\end{eqnarray}
It is worth mentioning that such type of terms, say,
$P^{|z_+^\bot,x_+\rangle}_{noEve}$, which  include such  elements
$P_{B_1|\chi_1^\bot}P_{B_2|\chi_2}$, is implicitly covered  in
(\ref{1d}). Precisely,  one can prove, e.g.,
$P^{|x_+,z_+\rangle}_{noEve}=P^{|x_-^\bot,z_+\rangle}_{noEve}$. We
plot the relation (\ref{1d})  in Fig. 1.
 Based on this
figure one  note $0.25 \leq P\preceq 0.28$. The case $ P=0.25$
occurs when Eve's
 projective basis reduces to one of the set given by (1) (see Fig. 1). In
 this respect, all elements in the right hand side of (\ref{1d}) tend to $1/2$, i.e.
$ P^{|i,j\rangle}_{noEve}=1/2, \forall i,j\epsilon
 \{x,y,z\}$ basis. This fact can be analytically checked.
Additionally, one can easily realise that the  maximum value
$P\simeq 0.28$ occurs when Eve's basis is different from those
used in the protocol, e.g., $(\theta,\phi)=(\pi/2,\pi/4)$. In
other words, when Eve measures the two qubits in the same basis
she will be detected at least by $72\%$, however, in [1] it was
approximately $50\%$.

\begin{figure}
   \includegraphics[width=.60\linewidth]{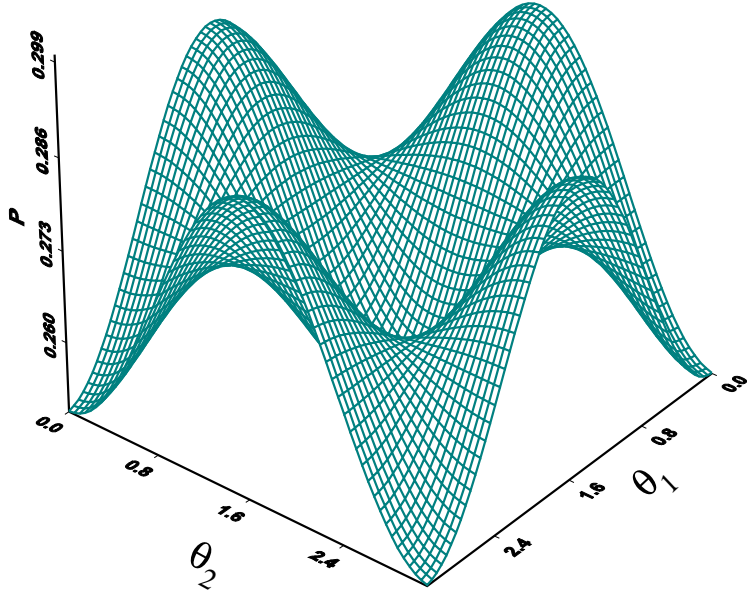}
  \caption{ The average probability of not detecting  Eve  during
the control mode of the  IRA. The angles $\theta_1$ and $\theta_2$
determine  Eve's projective measurement direction and $\phi_j=0$.
The angles are measured in radian.}
\end{figure}
Next, we draw the attention to  the second case, in which Eve
measures the two travelling qubits in two different bases, i.e.,
$j=1,2$ in (\ref{1cref}).  Consequently,  Eve  could obtain more
information than that extracted from  the  single-basis case.  In
this case the order of the travelling qubits should be taken into
account, where $P^{|z_+,x_+\rangle}_{noEve}\neq
P^{|x_+,z_+\rangle}_{noEve}$. Thus,  the expression (\ref{1d})
includes $24$ terms.
  Information about this situation  is
depicted in Fig. 2 for $\varphi_j=0$. From this figure it is
obvious that there is an improvement in Eve's information but not
too much, where $0.25 \leq P\preceq 0.3$. This means that the
probability of detecting Eve is at least $70\%$. The maximum value
$ P=0.3$ (minimum value $ P=0.25$)  occurs when Eve's bases reduce
to pair (one) of the $x,z,y$
 bases.  The natural question is: Can
Eve increase her information about the key by using three
different bases as those used in the protocol?  According to the
scenario described above, the answer is no since she, in this
case, may overcome difficulties in choosing which appropriate
projective bases for performing the measurement. This may decrease
her  information compared to those obtained from the previous
cases. To explain this point, assume
  that Bob sends two qubits in two
different bases to Alice. According to above strategy,  the
probability that Eve measures them in the  correct form in the
forward as well as in the backward paths  is $1/12$, i.e. in a
complete round the probability is $1/144\simeq 0.007$. Finally, we
should stress that the IRA cannot allow Eve to ascertain Alice
encoding with certainty, regardless of the choice of the
projection axis. This means that  Alice-Eve mutual information
after the IRA is less than unity. The reason is that Eve in the
IRA is going to use one or two or three projective bases. In each
case, as we discussed above, Eve cannot obtain full information
about the message. This fact will be elaborated  when we discuss
the 2CNOTA below.

We conclude this part by shedding some light on the eavesdropping
success probability as a function of $P$. Assume that $c$ is the
probability
 of occurrence  the control run. Thus the effective transmission rate reads $r=1-c$, which is
equal to the probability for a message transfer \cite{kim,Marco}.
The average probability of detecting Eve during a control run
equals $ d=1-P$, where $P$ is given by (\ref{1d}). Thus, the
probability that Eve acquires $n$ bits of full information without
being detected is \cite{kim}:

\begin{equation}\label{refereply}
    P_n(c,d)=\left[\frac{1-c}{1-c(1-d)}\right]^n.
\end{equation}
We apply the formula (\ref{refereply}) for the extreme values in
Fig. 1, i.e. $P=0.25$ and $0.28$, with $c=1/2$.  In this case, Eve
has correspondingly  probabilities  of about $1.1\%$ and $1.3\%$
to successfully eavesdrop $1$ byte (i.e., $8$ bits) of
information. It is evident that the  probabilities are  very low.

Now we draw the attention to the 2CNOTA. In this attack Eve
performs a first CNOT \cite{Nielsen} between the photons in
transit from Bob to Alice (control qubits) and her ancillae
(target qubits). The second CNOT is performed in the backward path
in the same scenario. This attack can give Eve full information on
the key for some protocols, e.g. \cite{Bennett}. In [1] it has
been argued that Eve can obtain full information about the 6DP
using this attack. This is obvious where the mutual entropy
between Alice and Eve as well as between Bob and Eve are
 unity (see (10) in [1]). On the contrary, here we
prove that the 2CNOT is not an efficient attack against the 6DP.
  This can simply be realized as
follows: Assuming that Eve uses the ancilla $|0\rangle_E$ with one
of the Bob qubits in the forward path. Then in the backward path,
Eve  measures her ancilla. It can be either in $|0\rangle_E$ or in
$|1\rangle_E$. This does not give her definite
 information about the encoded bits, where Alice could
execute one of four operations. More illustratively, when Eve
obtains $|0\rangle_E (|1\rangle_E)$ this means that Alice acted on
the qubits by one of the operators $\hat{I}$ or $\hat{Z}$
($\hat{X}$ or $\hat{Y}$). This point was not  discussed in [1].
Next, we extend  this conclusion when Eve uses two ancillae, say,
$|a\rangle_1^E|b\rangle_2^E$ where $a,b=0,1$. According to the
notion of the 6DP, Bob has to prepare two qubits from two
different bases as:
\begin{equation}\label{8}
   |\psi\rangle= |\psi_1\rangle\otimes|\psi_2\rangle=
   (\alpha_1|0\rangle_1+\beta_1|1\rangle_1)\otimes(\alpha_2|0\rangle_2+\beta_2|1\rangle_2),
\end{equation}
where $\alpha_j, \beta_j$ are c-numbers, which can be arbitrarily
chosen to provide the different forms of the states (1) and
$|\alpha_j|^2+|\beta_j|^2=1$. Eve executes her ancillae
$|a\rangle_1^{E}|b\rangle_2^{E}$ with the Bob qubits and performs
the first CNOT as:

\begin{equation}\label{9}
U_{cnot}(|\psi\rangle)|a\rangle_1^{E}|b\rangle_2^{E}=|\Psi_1\rangle=
(\alpha_1|0\rangle_1|a\rangle_1^{E}+\beta_1|1\rangle_1|1\oplus
a\rangle_1^{E})\otimes(\alpha_2|0\rangle_2|b\rangle_2^{E}+
\beta_2|1\rangle_2|1\oplus b\rangle_2^{E}).
\end{equation}
 Alice can act by any one of the four operators given above, say, $\hat{L}_j$, to encode both the qubits with the value $j=0$ or $1$.
  This could change the forms of the
 coefficients $\alpha_j,\beta_j$ to, say,  $\alpha'_j,\beta'_j$. The values of  $\alpha'_j,\beta'_j$ depend
 on which operator  Alice used in the encoding process. Then we
obtain:

\begin{equation}\label{10}
\hat{L}_j|\Psi_1\rangle=|\Psi_2\rangle=
(\alpha'_1|j\rangle_1|a\rangle_1^{E}+\beta'_1|1\oplus
j\rangle_1|1\oplus
a\rangle_1^{E})\otimes(\alpha'_2|j\rangle_2|b\rangle_2^{E}+
\beta'_2|1\oplus j\rangle_2|1\oplus b\rangle_2^{E}).
\end{equation}
Upon  returning on the backward path Eve performs the second CNOT
as:
\begin{equation}\label{11}
U_{cnot}|\Psi_2\rangle= (\alpha'_1|j\rangle_1+\beta'_1|1\oplus
j\rangle_1)\otimes(\alpha'_2|j\rangle_2+ \beta'_2|1\oplus
j\rangle_2)|j\oplus a\rangle_1^{E}|j\oplus b\rangle_2^{E}.
\end{equation}
 From (\ref{11}) it is obvious that the
ancillae have been disentangled from the qubits and they are
insensitive of the initial states. Next, Eve measures her ancillae
in the $z$ basis to get some information.  Actually, Eve's
information is restricted to her ancillae since she cannot access
the travelling qubits to see if the initial qubits are flipped or
not. This would confuse Eve since different operations can give
her the same results, as we mentioned above. For instance, assume
$a,b=0,0$ and Alice executed one of the operators $\hat{I}$ or
$\hat{Z}$ ($\hat{X}$ or $i\hat{Y}$). Thus Eve measurement outcome
is  $|0\rangle_1^{E}|0\rangle_2^{E}$
($|1\rangle_1^{E}|1\rangle_2^{E}$).  Similar conclusions can be
quoted for the other values of $a,b$. Furthermore, these results
are still valid even when Eve's ancillae comprise different bases,
e.g., one of the ancillae in $z$ basis and the other in $x$ basis.
We have checked this fact. According to this discussion
  Eve has a chance $50\%$ to get
the correct result and hence the mutual entropy between different
partners might  be written as: 
\begin{equation}\label{12}
I^{AE}_{2CNOT}=\frac{1}{2}, \quad I^{BE}_{2CNOT}=\frac{1}{2},
\quad I^{AB}_{2CNOT}=1.
    \end{equation}
The origin in  the last relation in (\ref{12}) is  that the 2CNOTA
does not perturb the message  and hence Alice and Bob mutual
information amounts always to 1.  In other words, in this attack
Eve's actions do not affect the users outcomes. Comparison between
the relation (\ref{12}) given above with Eq. (10) in [1] is
instructive.

It is worth mentioning that the analysis of the control sessions
for the 2CNOTA  is equivalent to that of using the identity
operator $\hat{I}$ in the encoding process by Alice in the above
analysis. In this case, Eve's anncillae do not change after a
whole run. This indicates that if the legitimate users restrict
the encoding operators to $\hat{I}$ and $\hat{Z}$ (instead of
four), Eve will obtain no information about the protocol from the
2CNOTA. In this regard, the protocol is secure against this type
of attack, however, it may be vulnerable against the others.

In conclusion,  the key extraction in the 6DP is based on  two
qubits from different bases. Considering  this property we treated
the IRA and 2CNOTA in this Comment. The rate of security obtained
from this treatment is higher than that shown in [1], in which the
security analysis was evaluated  to one-qubit case only.

\section*{References}

\end{document}